\documentclass[twoside]{LCWS11}
\usepackage[latin1]{inputenc}
\usepackage[dvips]{graphicx,epsfig,color}
\usepackage{wrapfig,rotating}
\usepackage{amssymb,amsmath,array}
\usepackage{cite}
\begin{document}
\title{
Prototyping of the ILC Baseline Positron Target} 
\author{Jeff Gronberg
\thanks{This work performed under the auspices of the U.S. Department of Energy by the Lawrence Livermore National Laboratory under Contract DE-AC52-07NA27344}, Craig Brooksby, Tom Piggott, Ryan Abbott, Jay Javedani, Ed Cook
\vspace{.3cm}\\
Lawrence Livermore National Laboratory\\
L-050, 7000 East Ave. Livermore, CA, 94550 - USA
}

\maketitle

\begin{abstract}
The ILC positron system uses novel helical undulators to create a powerful
photon beam from the main electron beam.  This beam is passed through a
titanium target to convert it into electron-positron pairs.  The target 
is constructed as a 1~m diameter wheel spinning at 2000~RPM to smear the
1~ms ILC pulse train over 10~cm.  A pulsed flux concentrating magnet is
used to increase the positron capture efficiency.  It is cooled to liquid
nitrogen temperatures to maximize the flatness of the magnetic field over
the 1~ms ILC pulse train.  We report on prototyping effort on this system.  
\end{abstract}

\section{Positron Source Overview}

\begin{figure}[bth]
\centerline{\includegraphics[width=\columnwidth]{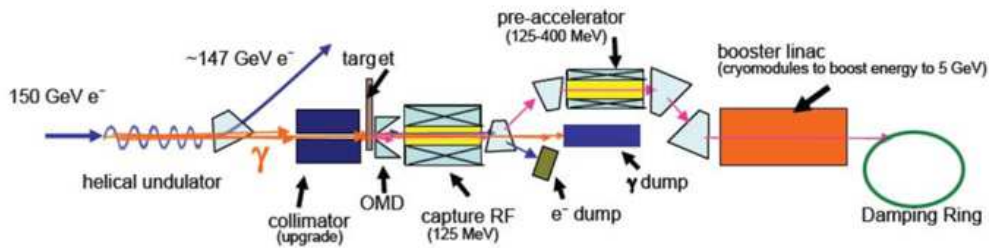}}
\caption{A schematic layout of the ILC positron source.}\label{Fig:Positron_layout}
\end{figure}

\begin{figure}
\centerline{\includegraphics[width=\columnwidth]{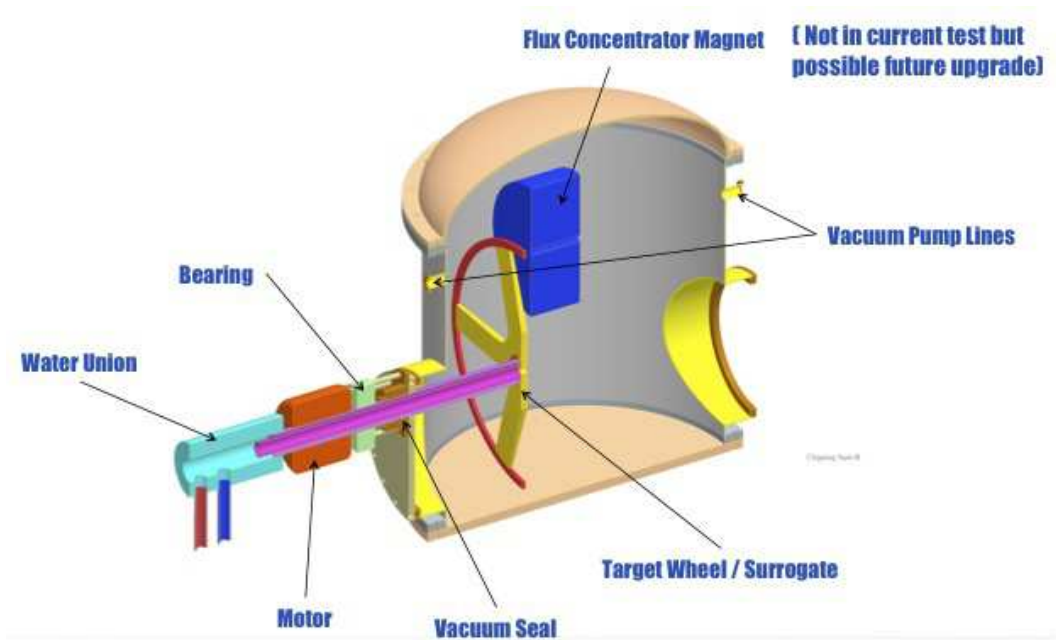}}
\caption{A schematic of the rotating target ferro-fluidic seal test stand.
}\label{Fig:tank_layout}
\end{figure}

The ILC positron source~\cite{ILCRDRvol3} will be required 
to generate two orders of magnitude
more positrons per second than any previous accelerator.  As shown in 
Figure~\ref{Fig:Positron_layout}, the baseline 
positron system envisions passing the main ILC electron beam through several
hundred meters of helical undulators in order to create a pulse of photons
with energies in the 10's of MeV and over 100~kW of average beam power.  The
photons must then be passed through a target in order to convert a fraction
of them into electron-positron pairs.  This target must operate in a unique
phase space compared to other target systems that have been fielded in the 
past.  The average power that the target must dissipate is low compared to
other systems but the power is concentrated into a small spot size and is 
deposited in a 1~ms time scale.  The energy deposition in a stationary target
would induce a stress in the target material which would exceed yield strength
of the material and would fracture the target.  The 1~ms timescale of the 
energy deposition makes it difficult to use motion of the target to 
distribute the energy deposition over a larger area.  A target moving at 
100~m/s will spread the energy deposition over a 10~cm stripe.  In order
to achieve this speed we have developed a target concept of a rotating
titanium wheel that has a diameter of 1~m and rotates at 2000~RPM, as shown
in Figure~\ref{Fig:tank_layout}.  

Cooling water flows through a double-walled
shaft to the target where it flows out each spoke, through a section of the
outer wheel, and back down a spoke.  The intensity of the electron-positron
beam emerging from the target makes the creation of a vacuum window to
separate the target volume from the subsequent accelerator sections
difficult.  No design has been found which can prevent such a vacuum window 
from melting during operation.  Therefore the target volume will share the
same vacuum as the subsequent capture accelerator sections in the positron
source system.  In order to have the rotating target in the same vacuum as
the accelerator we will need to have a rotating vacuum seal for the shaft.
Rotating vacuum seals based on ferrofluids exist and are available from
a number of vendors.  A fluid with suspended magnetic particles exists in
a gap between two counter rotating sets of permanent magnets.  This forms
the vacuum seal.  This solution must be prototyped and studied so that the
out-gassing rate from the ferro-fluid can be measured to see if a solution
for the vacuum pumping can be achieved.  The ability of the seal to perform
continuously for the planned 9 months of operation must also be determined.

A design for a pulsed flux concentrator has also been under study.  These
types of devices have been used before but usually with much shorter pulse
lengths.  This device will need to maintain a 1~ms flat top field during
the ILC bunch train.  A previous device~\cite{Brechna1965} 
created at SLAC during the 1960's
for a hyperon experiment, which was designed for a 40~ms pulse, was used
as the basis for the ILC design.  As well as maintaining a flat top during 
the ILC bunch train the device will need to be able to handle the 
radiation environment near the target.

\section{Prototyping of the Ferrofluidic Seal}

\begin{wrapfigure}{r}{0.5\columnwidth}
\centerline{\includegraphics[width=0.45\columnwidth]{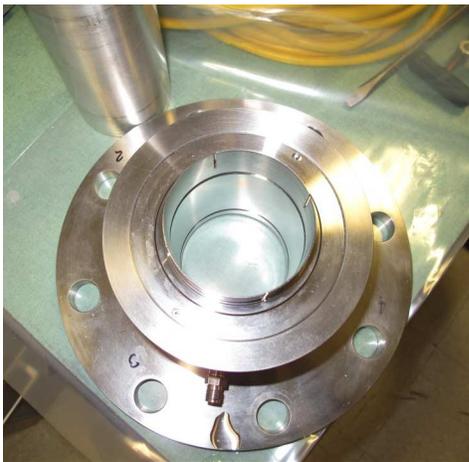}}
\caption{The RIGAKU ferrofluidic seal.  The inner bore of three inches 
diameter allows the central shaft to penetrate the vacuum.}
\label{Fig:Rigaku}
\end{wrapfigure}

Figure~\ref{Fig:Rigaku} shows a ferro-fluidic seal purchased from Rigaku
corporation.  The outer ring is stationary and sealed to the vacuum chamber.
The inner ring has a 3~inch inner bore through which the shaft can be mounted.
The ferro-fluid exists in the gaps between the two rings.   

\begin{wrapfigure}{r}{0.5\columnwidth}
\centerline{\includegraphics[width=0.45\columnwidth]{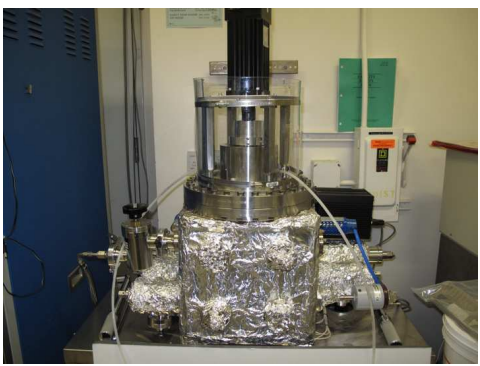}}
\caption{A test stand to do out-gassing studies of the ferro-fluidic seal while rotating at 2000~RPM.}\label{Fig:kishiyama}
\end{wrapfigure}

In order to test the out-gassing of the ferro-fluid into vacuum we modified
an existing out-gassing test system to be able to mount the seal and rotate
it at 2000~RPM.  Initial commissioning of the system had problems rotating the
seal at the full velocity, the drive motor would keep tripping off.  
Modifications were made to increase the available torque to drive the motor.
The devices have a choice of the type of ferro-fluid that is used and what
type of permanent magnet.  We initially chose a seal with the more
radiation hard permanent magnets and a more viscous ferro-fluid which should
reduce out-gassing.  However, a more viscous fluid also increases the torque
in the system and thus the energy deposited in the ferro-fluid.  While it was
rated to be able to run at 2000~RPM the Rigaku sealed failed after about
15~minutes of running at 2000~RPM.  It is believed that this
is a heating effect and the seal was returned to Rigaku for post-mortem 
analysis.  A second plug-compatible seal was sourced from FerroTec with a 
reduced viscosity ferro-fluid for testing.

\begin{figure}
\centerline{\includegraphics[width=\columnwidth]{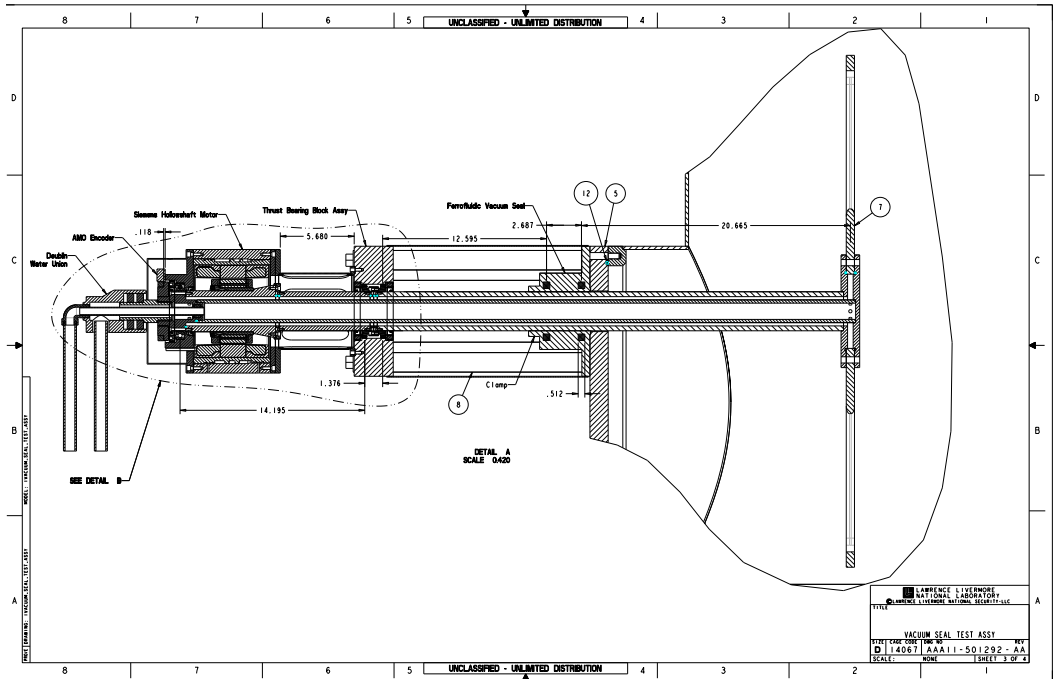}}
\caption{Assembly drawings for the target shaft assembly.
}\label{Fig:assembly}
\end{figure}

In parallel, construction of the full rotating shaft test stand is underway.
The detailed design drawings for the shaft are shown in 
Figure~\ref{Fig:assembly}.  The shaft is composed of two concentric pipes
to allow cooling water to flow to and from the shaft.  A rotating water
union is attached to the end of the pipe to mate with the outside water 
supply.  We plan to use the prototype titanium wheel that was created by the 
University of Liverpool for eddy current testing at the Daresbury lab.
Since it was not created with cooling channels the cooling water will 
only flow down the pipe and back.  The ferro-fluidic seal is mounted on the
bulkhead of the vacuum tank that we are using for this test.  Farther down
the shaft is a bearing block to provide support for the shaft and target.
A Siemens hollow shaft motor completes the assembly and will rotate the
shaft at 2000~RPM.

\begin{wrapfigure}{r}{0.5\columnwidth}
\centerline{\includegraphics[width=0.45\columnwidth]{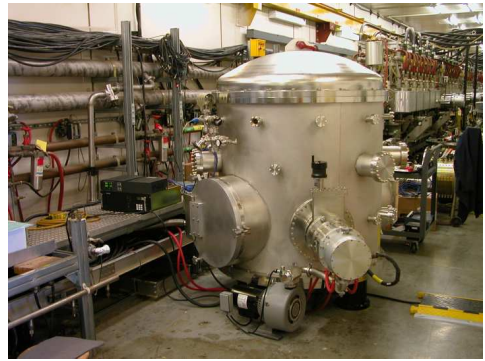}}
\caption{The vacuum tank at LLNL being used for the rotating seal test.}
\label{Fig:tank}
\end{wrapfigure}

As of LCWS11 we had commissioned a vacuum tank as shown in 
Figure~\ref{Fig:tank} and had received all of the manufactured parts for
assembly of the shaft.  Commissioning of a data acquisition and slow
control system was ongoing.  The long term testing will continually monitor
cooling water flow rates to the ferro-fluidic seal and Siemens motor.  
Temperature monitors will be in place on the ferro-fluid seal, bearing
block and motor as well as the cooling water.  Three-axis vibration
sensors will be mounted on the ferro-fluid seal, bearing block and motor.
We will use this to monitor any degradation in the bearings over time.

\clearpage

\section{Prototyping of the Pulsed Flux Concentrator Magnet}

\begin{figure}
\centerline{\includegraphics[width=0.8\columnwidth]{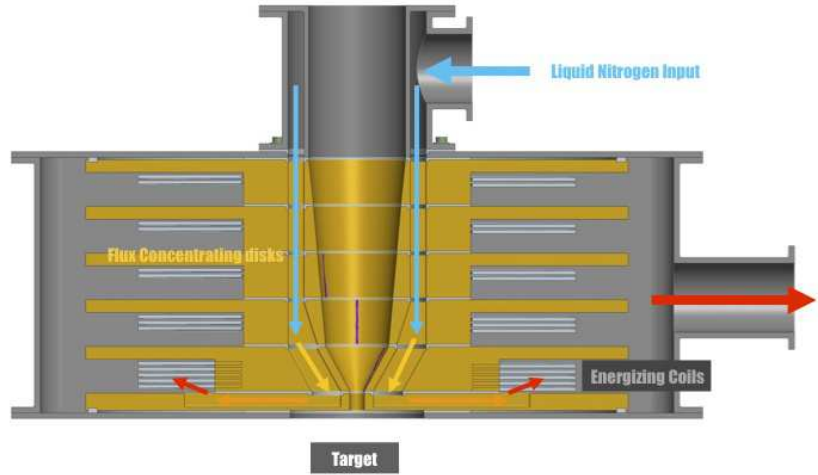}}
\caption{Schematic of the pulsed flux concentrator.  Energizing coils 
induce a current in the concentrating plates to create the magnetic field 
in the bore.  The entire assembly is bathed in liquid nitrogen to reduce
the electrical resistance of the plates and help maintain the current over
the 1~ms ILC bunch train.}\label{Fig:pfc_layout}
\end{figure}

\begin{wrapfigure}{r}{0.5\columnwidth}
\centerline{\includegraphics[width=0.45\columnwidth]{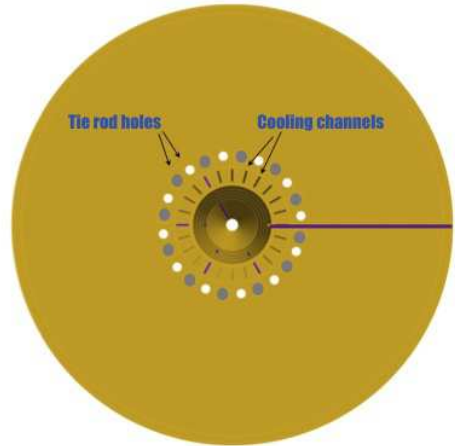}}
\caption{One of the flux concentrating disks.  An insulating layer forces
the induced current to flow around the bore.}\label{Fig:pfc_disk}
\end{wrapfigure}

A pulsed flux concentrator (PFC) is basically a transformer where a set of 
energizing coils induces a current in a set of concentrating plates.
Figure~\ref{Fig:pfc_layout} shows the six concentrating plates interleaved
with the five energizing coils.  Figure~\ref{Fig:pfc_disk} shows an example
of one of the concentrating plates.  The insulating gap that runs from
the bore to the edge is critical to successful operation.
Without a gap the induced currents in the plate would have the opposite 
sense as the currents in the coils and would cancel
the magnetic field generated from the coils.  The insulating
gap from the bore to the outer edge prevents current from circulating around
the outside edge and forces the current to travel around the bore in the
same sense as the coils.  This has the effect of concentrating the field
from the coils into a smaller cross-sectional area and creating a higher
field in the bore. One of the limitations of this technique is that currents
in the plate will dissipate over time due to ohmic resistance losses in 
the plate reducing the field.
On a long enough timescale the currents in the concentrating disk will
fall to zero and the field will return what would be induced by just the 
energizing coils.  For this application we would like to have a reasonably
constant field over the 1~ms pulse length of the ILC bunch train.  Therefore
we plan to use OFHC copper and cool it in a liquid nitrogen bath in order
to decrease it's resistance and increase the usable length of the pulse.

The site of highest energy deposition from operation of the device is 
in the concentrating plates near
the bore of the magnet where the currents are highest.  The first concentrating
plate dissipates about 300~W and the second concentrating plate dissipates
about 500~W at full 5~Hz operation.  In order to provide the best cooling
in this region a set of cooling channels are designed to run along the bore
from the back end of the device to the front and then discharge into the 
larger liquid nitrogen bath.  The cooling channels are designed as slits that 
are positioned radially around the bore as shown in Figure~\ref{Fig:pfc_disk}.
  This design requires an insulating
seal between the concentrating plates to allow the cooling fluid to flow
between the concentrating plates and to create a vacuum seal between 
the magnet bore vacuum and the
liquid nitrogen bath.  This is an extreme technical challenge as the seal
must be electrically insulating, radiation hard, and able to withstand
repetitive impacts from the 5~Hz pulsed operation.  We have created a concept
for a seal based on flexible graphite to create the radiation hard seal
with a layer of Zirconia Toughed Alumina to provide the radiation hard 
electrical connection.  Initial prototyping test will be carried out to
determine if this is a viable solution.

\begin{wrapfigure}{r}{0.5\columnwidth}
\centerline{\includegraphics[width=0.45\columnwidth]{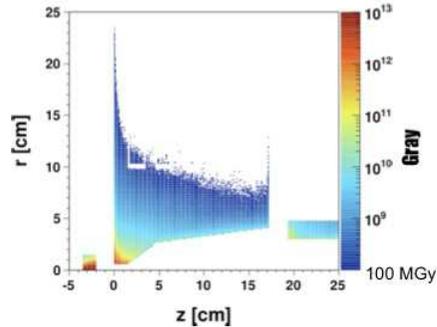}}
\caption{A Monte Carlo simulation of radiation deposition in the material
of the pulsed flux concentrator from DESY/Zeuthen over 9 months
of ILC running.  The 100~MGy line is the limit for using organic insulators.
All insulation of the plates at higher radiation levels must use ceramics.}
\label{Fig:ushakov_radiation}
\end{wrapfigure}

The creation of a liquid tight cavity for the nitrogen bath also requires 
that the gap in the concentrating plate be filled with a radiation hard
electrical insulator that can be bonded into place to form a liquid tight
seal.  This bond will be a site of high stress during the pulse.  The 
requirement that this bond survive 100~million pulses over the 9 months
of operation 
leads to a design choice of Zirconia Toughed Alumina.  This material
should be able to survive the repetitive stress.  
This may be difficult to manufacture
and we are searching for vendors who can do this type of braze.

The device sits a few centimeters from the target and is exposed
to the radiation field from the device.  Electrons, positrons and photons
will emerge from the target and hit the device.  
Figure~\ref{Fig:ushakov_radiation}~\cite{Andriy} 
shows the expected dose in the device
for 9 months of operation as a function of position.  At the front face of
the device, which is directly exposed to the charged particle flux, the dose
can be TeraGray.  The solid copper material of the disks provides
for self-shielding of the charged particles and dose falls off rapidly.
However the photon flux from the target is harder to shield against.  Photon
conversions will provide a radiation flux at all distances falling off 
as distance squared and from the shielding effect of material in the path
of the photon.  The energizing coils and concentrating plates will need 
to be electrically insulated from each other.  
The organic insulator with the best radiation survival
is Kapton which is rated out to 100~MGray.  
From Figure~\ref{Fig:ushakov_radiation}
we can see that it may be usable as an insulating material on the energizing
coils but is unusable anywhere closer to the bore of the device.

\section{Future Work}

As of LCWS11 the out-gassing test stand for the ferro-fluidic seal was 
being commissioned and the full rotating shaft test stand was 
under construction.  Assembly drawings for the pulsed flux concentrator were
being started and a set of prototype tests of the seals for the liquid
nitrogen containment were planned.  Once component testing is complete
we will proceed to create a pulser to drive
a prototype magnet at full current but reduced repetition rate and create
a prototype magnet with the important features that will be required for 
realization of a final device at ILC.

\section{Acknowledgments}
The author is grateful for the assistance of the ILC positron group in
these efforts.
Particularly the DESY/Zeuthen group, Sabine Riemann, Andriy Ushakov and
Friedrich Staufenbiel for their calculations of radiation damage in the
pulsed flux concentrator and the ANL group, Wei Gai and Wanming Liu for
calculations of the capture efficiency of the pulsed flux concentrator.
Thanks to Ian Bailey for the loan of the prototype titanium target wheel.


\begin{footnotesize}


\end{footnotesize}


\end{document}